\documentclass[sigconf]{acmart}

\usepackage{balance}

\def\BibTeX{{\rm B\kern-.05em{\sc i\kern-.025em b}\kern-.08emT\kern-.1667em\lower.7ex\hbox{E}\kern-.125emX}}
    
\copyrightyear{2020}
\acmYear{2020}
\setcopyright{acmlicensed}
\acmConference[SPAI '20] {1st Security and Privacy on Artificial Intelligent Workshop}{October 6, 2020}{Taipei, Taiwan}
\acmBooktitle{1st Security and Privacy on Artificial Intelligent Workshop (SPAI'20), October 6, 2020, Taipei, Taiwan}
\acmPrice{15.00}
\acmDOI{10.1145/3385003.3410921}
\acmISBN{978-1-4503-7611-2/20/10}

\DeclareMathOperator*{\kHz}{kHz}
\DeclareMathOperator*{\dB}{dB}
\DeclareMathOperator*{\ms}{m/s}
\DeclareMathOperator*{\ns}{ns}
\DeclareMathOperator*{\sneg}{s^{-1}}
\DeclareMathOperator*{\m}{m}
\DeclareMathOperator*{\cm}{cm}
\DeclareMathOperator*{\s}{s}

\begin{document}

\fancyhead{}

\title{Towards Resistant Audio Adversarial Examples}

\author{Tom D\"orr}
\authornote{Both authors contributed equally to this research.}
\email{tom.doerr@tum.de}
\author{Karla Markert}
\authornotemark[1]
\author{Nicolas M. M\"uller}
\author{Konstantin B\"ottinger}
\email{karla.markert, nicolas.mueller, konstantin.boettinger@aisec.fraunhofer.de}
\affiliation{%
  \institution{Fraunhofer AISEC}
  \streetaddress{Lichtenbergstr. 11}
  \city{Garching}
  \country{Germany}
  \postcode{85748}
}

%
\renewcommand{\shortauthors}{Dörr and Markert, et al.}

%
\begin{abstract}
Adversarial examples tremendously threaten the availability and integrity of machine learning-based systems.
While the feasibility of such attacks has been observed first in the domain of image processing, recent research shows that speech recognition is also susceptible to adversarial attacks.
However, reliably bridging the air gap (i.e., making the adversarial examples work when recorded via a microphone) has so far eluded researchers.

We find that due to flaws in the generation process, state-of-the-art adversarial example generation methods cause overfitting because of the binning operation in the target speech recognition system (e.g., Mozilla Deepspeech).
We devise an approach to mitigate this flaw and find that our method improves generation of adversarial examples with varying offsets. 
We confirm the significant improvement with our approach by empirical comparison of the edit distance in a realistic over-the-air setting.
Our approach states a significant step towards over-the-air attacks.
We publish the code and an applicable implementation of our approach.
\end{abstract}



\begin{teaserfigure}
  \includegraphics[width=\textwidth]{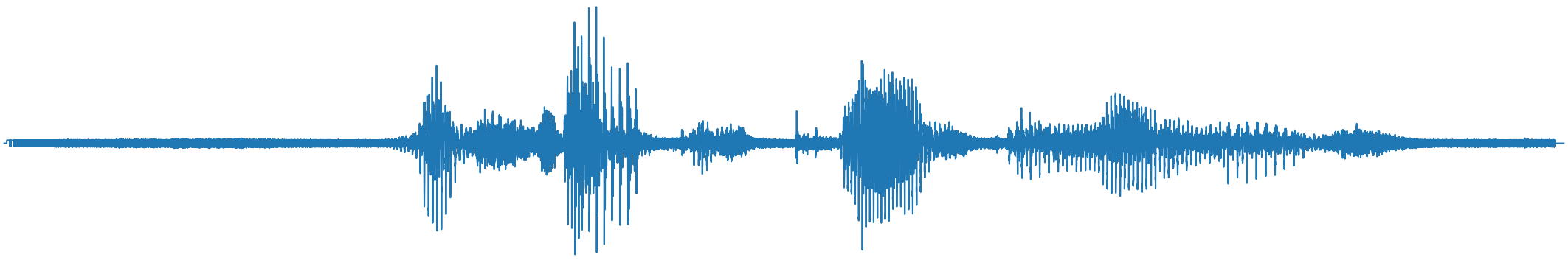}
  \caption{Offset-resistant audio adversarial example. \emph{Original phrase:} ``they're calling to us not to give up and to keep on fighting''. \emph{Target phrase:} ``even coming down on the train together she wrote me''.}
  \Description{Offset-resistant audio adversarial example. \emph{Original phrase:} ``they're calling to us not to give up and to keep on fighting''. \emph{Target phrase:} ``even coming down on the train together she wrote me''.}
  \label{fig:teaser}
\end{teaserfigure}

\maketitle

\section{Introduction}
People all over the world increasingly rely on voice assistants to accomplish their day to day tasks \cite{petrock2019us}. 
With this increase in usage, voice assistants get more and more access to personal data and control over IoT devices. 
Hence, exploiting flaws in such assistants through weaknesses in the speech recognition system becomes more attractive for criminals.
Thus, attacks on speech recognition models pose a threat to the users' privacy and the security of their connected devices.

One possible attack scenario is the generation of adversarial examples, first discussed in \cite{biggio2013evasion, szegedy2013intriguing}.
In this kind of attack, a slight distortion is added to the input, in our case the audio data.
This perturbation is nearly imperceptible to humans but causes the neural network to misclassify the input.
Put in our application scenario: a speech recognition algorithm is misled to detect a completely different text.
Humans, however, will still recognize the original text and might not even notice the adversarial signals. 

This discrepancy in the audio interpretation between humans and automatic speech recognition systems could cause unsuspecting end users to play manipulated audio and, for example, involuntarily order items online, send malicious messages, or share personal data.

A crucial step in this attack is bridging the air gap, that is to say, playing the adversarial sound file over speakers instead of directly feeding it digitally to the speech recognition system.
Making robust adversarial audio examples has eluded researchers: 
methods presented in \cite{carlini2018audio, qin2019imperceptible, yakura2018robust, schonherr2019robust} work well when examples are fed directly into the speech recognition system, but fail to reliably bridge the air gap.

In this paper, we present an important ingredient necessary in order to make adversarial examples resistant against the air gap:
resistance against varying offsets, i.e., small changes in the signal binning preprocessing step.
We motivate this insight by showing empirically that adversarial examples generated with state-of-the-art methods overfit to the binning during preprocessing.
This results in poor generalization in realistic environments, when noise and delay change the way the audio data is binned for the Mel Frequency Cepstral Coefficient (MFCC) computation.
Based on this insight, we introduce a modification to the algorithm presented in \cite{carlini2018audio}.
With this, we can produce adversarial examples that are invariant to pre-pended silence, when fed directly to the speech recognition system.
Further, even in the over-the-air setting their assigned transcriptions are significantly closer to the target labels, even without considering the room reverberation in the generation process.

Our approach is based on the code used in \cite{carlini2018audio} and can be generalized to more recent approaches.
All source code is open-sourced for full reproduceability\footnote{The code for this paper is published at \url{https://github.com/Fraunhofer-AISEC/towards_resistant_audio_adversarial_examples}.}.
We also provide pre-calculated adversarial examples for the interested reader.

The paper is structured as follows:
in Section~\ref{sec:relatedWork}, we provide a brief overview of the research on adversarial examples in the image and audio domain. 
In Section~\ref{sec:offsetEffect}, we evaluate the offset-resistance for two well-known approaches. 
Based on these insights, we classify and explain our approach in Section~\ref{sec:problemDefinition}. 
In Section~\ref{sec:empiricalEvaluation}, we provide an empirical analysis of our method, including an evaluation of audio files played in a realistic over-the-air setting. 
We summarize our approach in Section~\ref{sec:conclusion}.

\section{Related Work}\label{sec:relatedWork}
With the rise of neural networks, adversarial examples and their mitigation \cite{zhang2018adversarial,carlini2019evaluating, hu2019adversarial} have become subject to competitive research, especially in the image domain, see Section~\ref{ssec:advImages}.
Analog to images, one can generate adversarial examples for audio data, see Section~\ref{ssec:advAudio}.

The first adversarial examples to non-linear algorithms were generated by Biggio et al.~\cite{biggio2013evasion} and Szegedy et al.~\cite{szegedy2013intriguing} in 2013.
Up until now, various attacks on plenty of datasets focusing on images have been presented (for an overview, see, e.g., \cite{zhang2018adversarial}).
Later, adversarial audio started to get analysed as well \cite{yakura2018robust,qin2019imperceptible,carlini2018audio,taori2018targeted,khare2018adversarial,schonherr2019robust,du2019sirenattack}.

In the following, we shortly discuss adversarial examples for images to then introduce its application in automatic speech recognition.

\subsection{Adversarial Examples for Images}\label{ssec:advImages}

Research into adversarial images has covered many different application areas: e.g.~face recognition \cite{taigman2014deepface}, traffic sign recognition \cite{lu2017adversarial}, and object recognition \cite{szegedy2013intriguing}.

One aspect is also to find real-world adversarial examples: objects that are miss-classified when photographed \cite{athalye2017synthesizing} or filmed \cite{tencent2019experimental}.
Further, researchers have found non-manipulated photos that are miss-classified among different models \cite{hendrycks2019natural}.

Different mitigation methods and defenses against adversarial attacks have been discussed in \cite{madry2017towards, papernot2016distillation, liu2018towards}.
As pointed out in \cite{carlini2017adversarial}, however, many of them can be circumvented by an adversary who knows that a detection scheme was put in place.
Thus, there are only very few defense mechanisms that are actually considered helpful (mainly adversarial training) \cite{carlini2019lessons}.

Adversarial examples for audio data are somewhat different to create because of the time-dependency in psychoacoustic processing and the different resulting objective functions.

\subsection{Adversarial Examples for Audio Data}\label{ssec:advAudio}
Building on research in the image domain, Carlini and Wagner designed the first successful targeted adversarial attack in the audio domain \cite{carlini2018audio} and, thus, further improved previous work on induced misclassification in speech recognition systems \cite{carlini2016hidden, song2017inaudible, zhang2017dolphinattack}. 

Their attack uses Deepspeech v0.1.0 (a more current version\footnote{See \url{https://github.com/carlini/audio\_adversarial\_examples}, last checked July 17, 2020.} attacks v0.4.1), a speech recognition system developed by Hannun et al.~\cite{hannun2014deep}.
Deepspeech, like other speech recognition systems, pre-processes the audio data by grouping the raw audio signal into 50 buckets per second that may overlap \cite{carlini2018audio}. 
Then, via the Fast Fourier Transform, the frequency spectrum for each bucket is calculated.
This process is called Short-Time Fourier Transform.
Finally, the spectra are passed through Mel-Filters to derive a Mel-Spectrogram in order to calculate the Mel-Frequency Cepstral Coefficients (MFCC) \cite{hu2019adversarial}. 
By mapping the buckets to the MFCC, the audio is compressed into a representation that keeps the information most important to the human auditory system, which makes the MFCC suitable for speech recognition systems \cite{mermelstein1976distance, davis1980comparison, bridle1974experimental}.

Deepspeech is trained using the so-called Connectionist Temporal Classification (CTC), where the audio data is labelled with a transcribed sentence without segmentation, thus, without knowing the exact position of each word in the file \cite{carlini2018audio}.

As noted by the authors, the attack presented in \cite{carlini2018audio} does not work in a scenario where the adversarial example is played by a speaker and recorded by a microphone.

In accordance with the literature, we refer to this setting as \emph{over-the-air}.
Furthermore, we call adversarial examples that are also feasible under over-the-air conditions \emph{resistant adversarial examples}.
This notation differs from, e.g., \cite{yakura2018robust, schonherr2019robust} but makes the distinction between robust models (models that are not susceptible to adversarial examples) and resistant adversarial examples (examples that are able to bridge the air-gap) more clear.

Yakura and Sakuma, the authors of \cite{yakura2018robust}, include a band-pass filter, impulse response, and white Gaussian noise in the training process in order to tackle the following three challenges in the over-the-air setting: frequency range limitations of speakers and microphones, reverberations, and background noise.
A band-pass filter removes frequencies above and below a certain threshold to only allow frequencies that lie within a certain frequency band to pass.
For the simulation of reverberations, \cite{yakura2018robust} uses 615 recorded room impulse responses.
The experiments are also based on Deepspeech v0.1.1 as for the original code in \cite{carlini2018audio}.
As reported by the authors, their generated adversarial examples bridge the air-gap with a success rate of $50\% - 100\%$.
However, their example phrases are a) very short and b) mainly based on music recordings which cut through additional noise more easily than speech recordings.

A rather new approach was taken in \cite{qin2019imperceptible}, which includes psychoacoustic models in the training with the aim of better imitating human hearing in order to improve the imperceptibility.
Furthermore, the authors used an acoustic room simulator to create room impulse responses.
This process enhances clean audio with reverberations in an effort to make the result more resistant to the air gap.
The experiments are based on the speech recognition system Lingvo \cite{shen2019lingvo}.
Here, the MFCCs are fed to a sequence-to-sequence model consisting of convolutional and LSTM layers.

Similarly, \cite{schonherr2019robust} has included psychoacoustic perception based on MP3 and room simulation in order to generate resistant examples, reporting a success rate of up to $3.7\%$ in real over-the-air attacks.
Very short phrases that were successfully transmitted over-the-air can also be found in \cite{du2019sirenattack}.

A rather new attack that directly inserts commands to the microphone via laser, has been introduced in \cite{sugawara2020light}.
Side-channel-based scenarios are, however, not considered in this work.

There are only few approaches to mitigate attacks on speech recognition systems \cite{yang2018characterizing}.
One approach is to include adversarial examples in the training data, as has been proven successful for images.

In the following, we focus on \cite{carlini2018audio} (by Carlini and Wagner) and \cite{qin2019imperceptible} (by Qin et al.) and evaluate their algorithms' invariance to offset.
We have picked these two approaches, since both works include comprehensive online published code and, hence, enabled us to generate adversarial examples ourselves.

\section{Offset's Effect in Existing Methods}\label{sec:offsetEffect}
Prior work shows that applying rotations to images can change their algorithmically assigned label \cite{fawzi2015manitest,azulay2018deep}.
Altering the classification this way enables one to generate adversarial images without the need to change the rotated image itself. 
A transformation in the audio domain that is analogous to image rotations is the \emph{offset}, where silence of a certain duration is prepended to the audio file.
This way, the audio data is just slightly shifted, but remains unchanged in itself.

The reason why an offset influences Deepspeech's prediction lies in its architecture.
The offset changes the way the audio is being processed in Deepspeech. 
Since it uses buckets to group the audio samples for the preprocessing steps (see Section~\ref{ssec:advAudio}), a small shift may already lead to very different data per bucket, resulting in a different output from the preprocessing step.
Hence, by prepending silence, we get a different input to the neural network that Deepspeech uses for recognizing the speech. 
This property also holds for other speech recognition models that use a similar preprocessing technique (see Section~\ref{ssec:qin}).
In the next subsections, we evaluate the offset-resistance for the approaches provided in \cite{carlini2018audio} and \cite{qin2019imperceptible}.

An overview of the different settings used later in this paper is provided in Table~\ref{tab:settings}.

\begin{table}[h]
    \centering
    \caption{Different settings for testing offset-resistance}
    \begin{tabular}{cp{2.8cm}p{2.8cm}}
    \toprule
        Setting & Original Label & Target Label \\
    \midrule
        1 & ``follow the instructions here'' & ``the shop is closed on mondays''\\
        2 & ``they're calling to us not to give up and to keep on fighting'' & ``even coming down on the train together she wrote me''\\
        3 & ``the shop is closed on mondays'' & ``i'm going away he said''\\
        4 & ``even coming down on the train together she wrote me'' & ``it must have fallen while i was sitting over there''\\
        5 & music (``To The Sky'' by Owl City) & ``open the door''\\
    \bottomrule
    \end{tabular}
    \label{tab:settings}
\end{table}

\subsection{Effects for Carlini and Wagner}\label{ssec:carlini}
In the following, we analyse the effect of shifting offsets for adversarial examples created as proposed by \cite{carlini2018audio}.

\begin{figure}[h]
    \centering
    \includegraphics[width=0.45\textwidth]{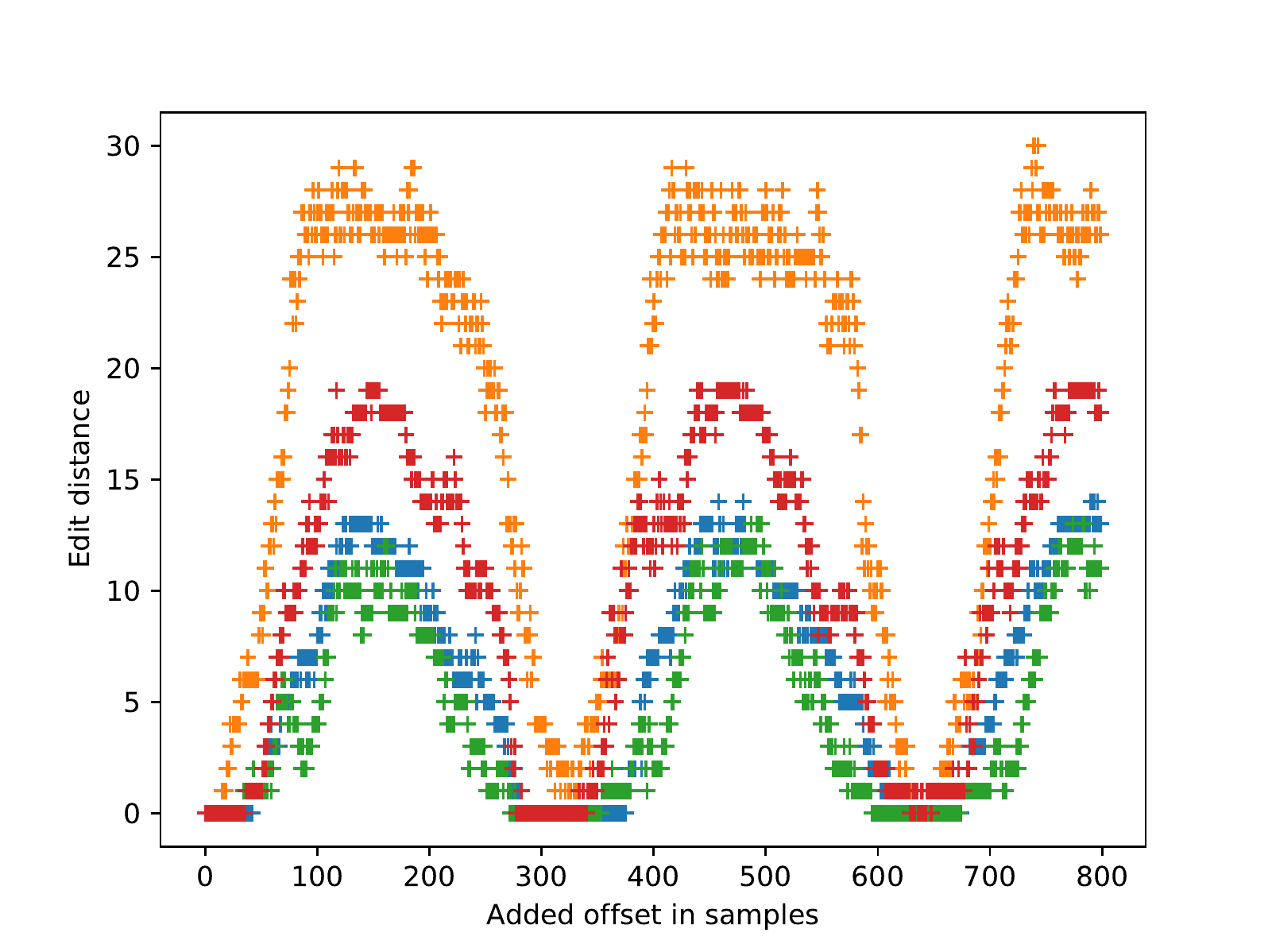}
    \caption{
        The plot shows the edit distance between prediction and targeted adversarial label for settings 1 - 4 from Table~\ref{tab:settings} ($y$ axis: edit distance; $x$ axis: added offset in samples). 
    }
    \label{fig:targetedCarlifiniOffsetVariation10Samples}
\end{figure}

Figure~\ref{fig:targetedCarlifiniOffsetVariation10Samples} shows how strongly adversarial examples trained via \cite{carlini2018audio} overfit to the precise binning operation.
We perform a simple experiment: given adversarial audio example found via \cite{carlini2018audio}, we prepend a very short sample of silent audio, i.e.~a null vector of length $R \in \left\{0, \, \ldots, \, 800\right\}$, corresponding to less than 0.05 seconds of silence.
We feed the result to Deepspeech and record the edit distance \cite{levenshtein1966binary} to the adversarial target phrase.
On the $x$ axis, we find $R$, the added offset measured in samples, while the edit distance to the target is displayed as a function of the offset.
We perform the experiment for four different audio adversarial examples (settings 1-4, as shown in Table~\ref{tab:settings}, phrases taken from \cite{mozilla2017common}), when directly fed into Deepspeech v0.4.1.
In order to facilitate comparison to \cite{qin2019imperceptible}, we have chosen random original label and target label combinations of similar length.

As one can see from Figure~\ref{fig:targetedCarlifiniOffsetVariation10Samples}, there is a minimum at offset zero, which corresponds to the adversarial audio without an offset.
There, the edit distance between the transcription and the target label is zero.
Hence, this attack is successful as constructed.

Adding some non-zero offset, however, reduces the effectiveness of the attack:
the edit distances increase to up to 30, rendering the attack ineffective.

This dependence on the offset can be explained as follows: the sampling rate of $16\kHz$, divided by the number of buckets per second, 50, leads to 320 samples per bucket.
Thus, a shift of 320 or 640 samples results in the same division into buckets as an offset of zero samples (no offset).

\begin{figure}[h!]
    \centering
    \includegraphics[width=0.22\textwidth]{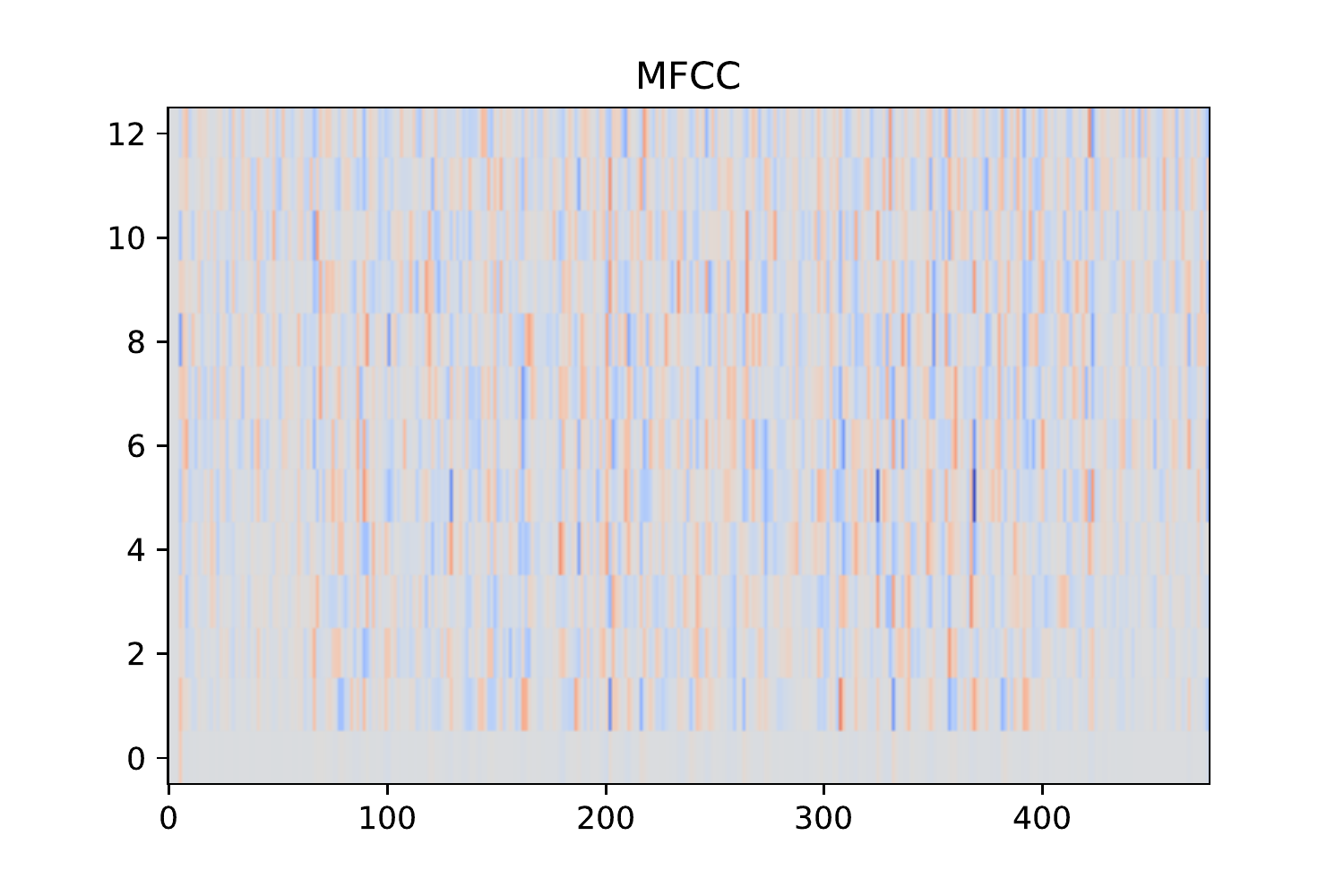}
    ~ 
    \includegraphics[width=0.22\textwidth]{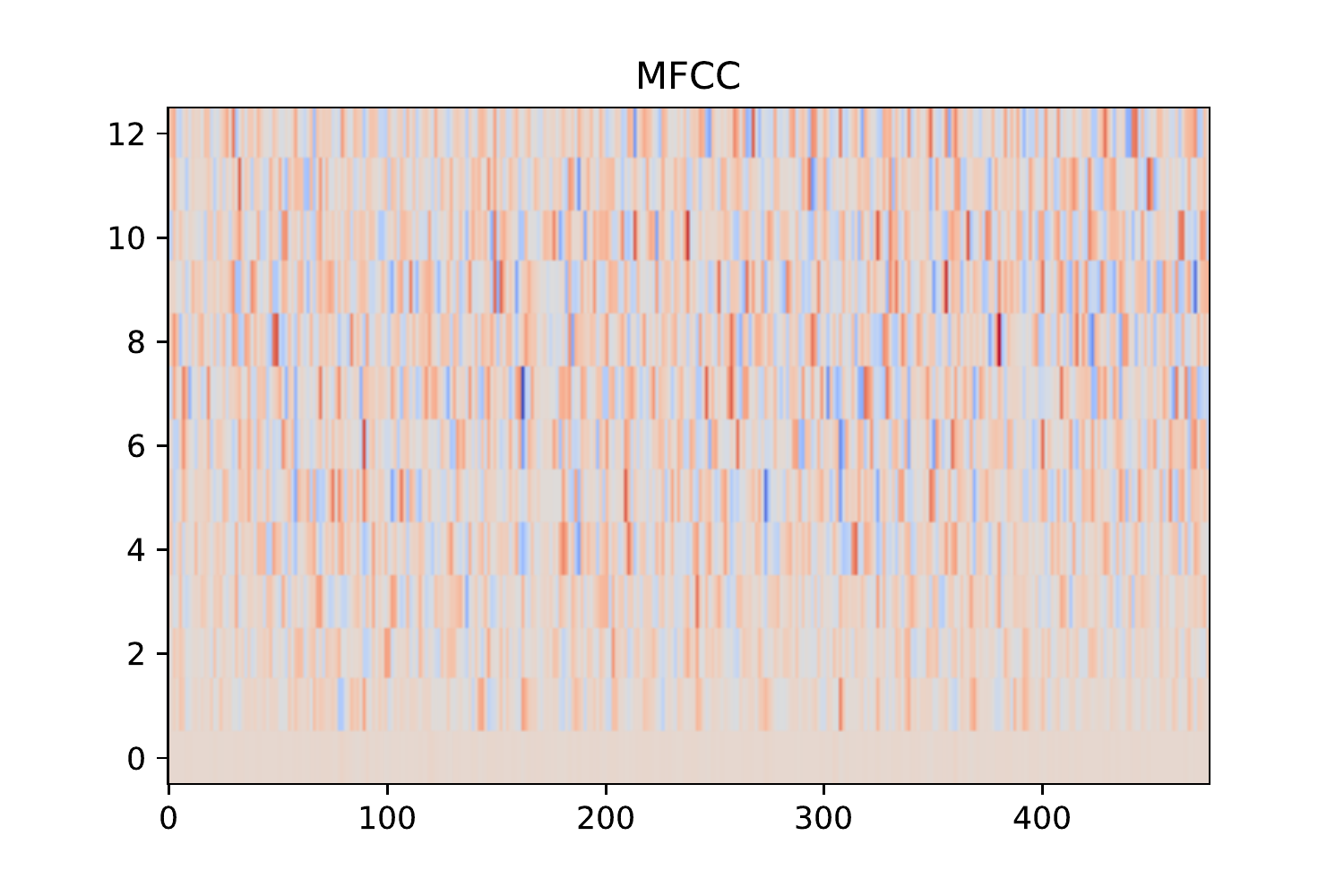}
    \caption{Difference between the MFCC output for offsets 0 and 160. Left, for the original audio and its shift; right, for adversarial audio and its shift ($y$ axis: frequency; $x$ axis: time window). The more intense the colors, the higher the absolute value. The adversarial audio is generated based on \cite{carlini2018audio}, setting 1, Table~\ref{tab:settings}.}
    \label{fig:differences}
\end{figure}

Next, we compare the MFCC without prepending silence with prepending a 160-samples long vector of zeros (i.e., silence).
The results are displayed in a spectrogram in Figure~\ref{fig:differences}, left.
As can be seen from the color spectrum, the absolute values are below those of the differences for the adversarial audio file with no prepended silence and the original audio file with 160 prepended samples of absolute silence, see Figure~\ref{fig:differences}, right.

\subsection{Effects for Qin et al.}\label{ssec:qin}
\begin{figure}[h]
    \centering
    \includegraphics[width=0.22\textwidth]{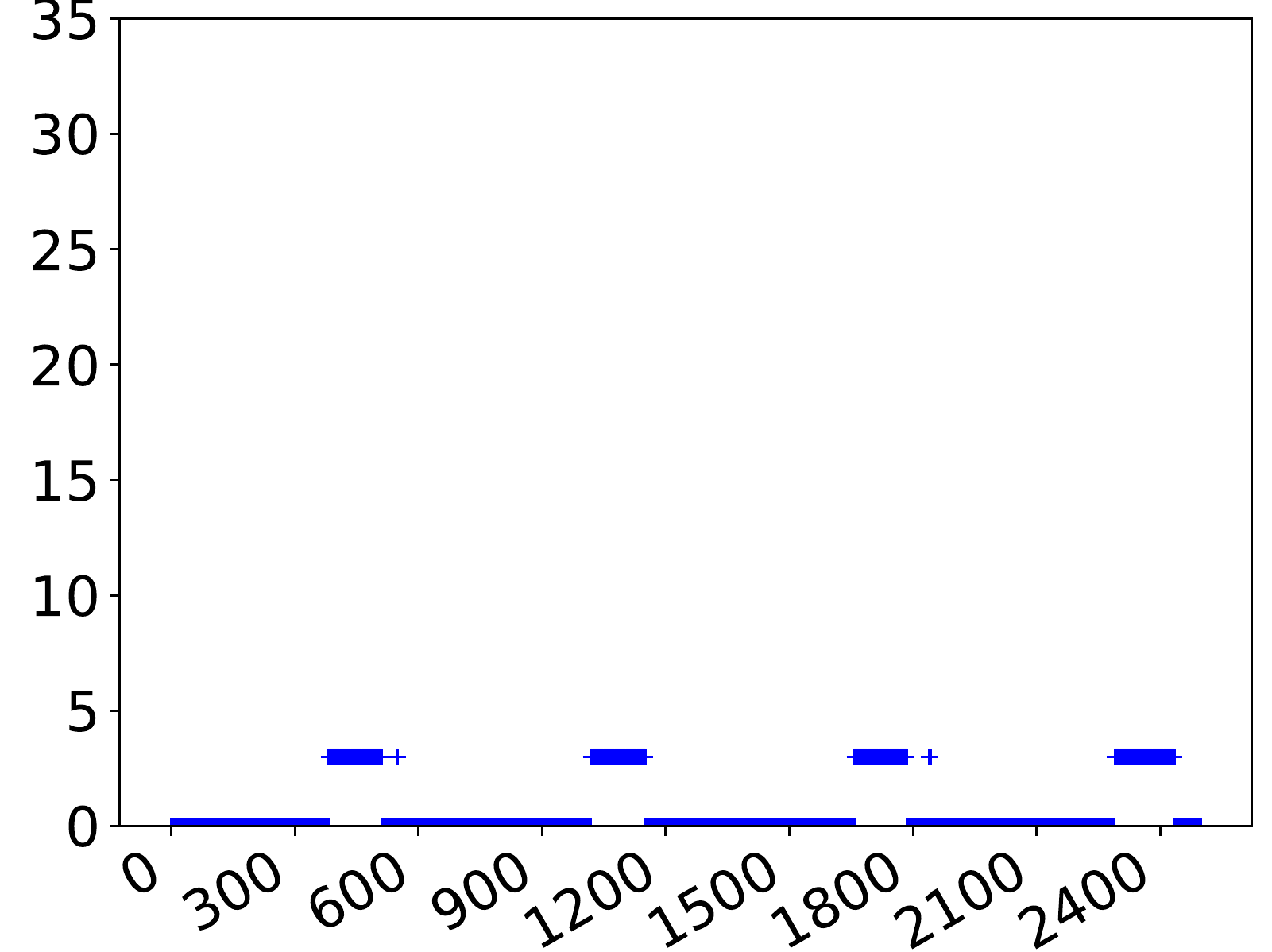}
    ~ 
    \includegraphics[width=0.22\textwidth]{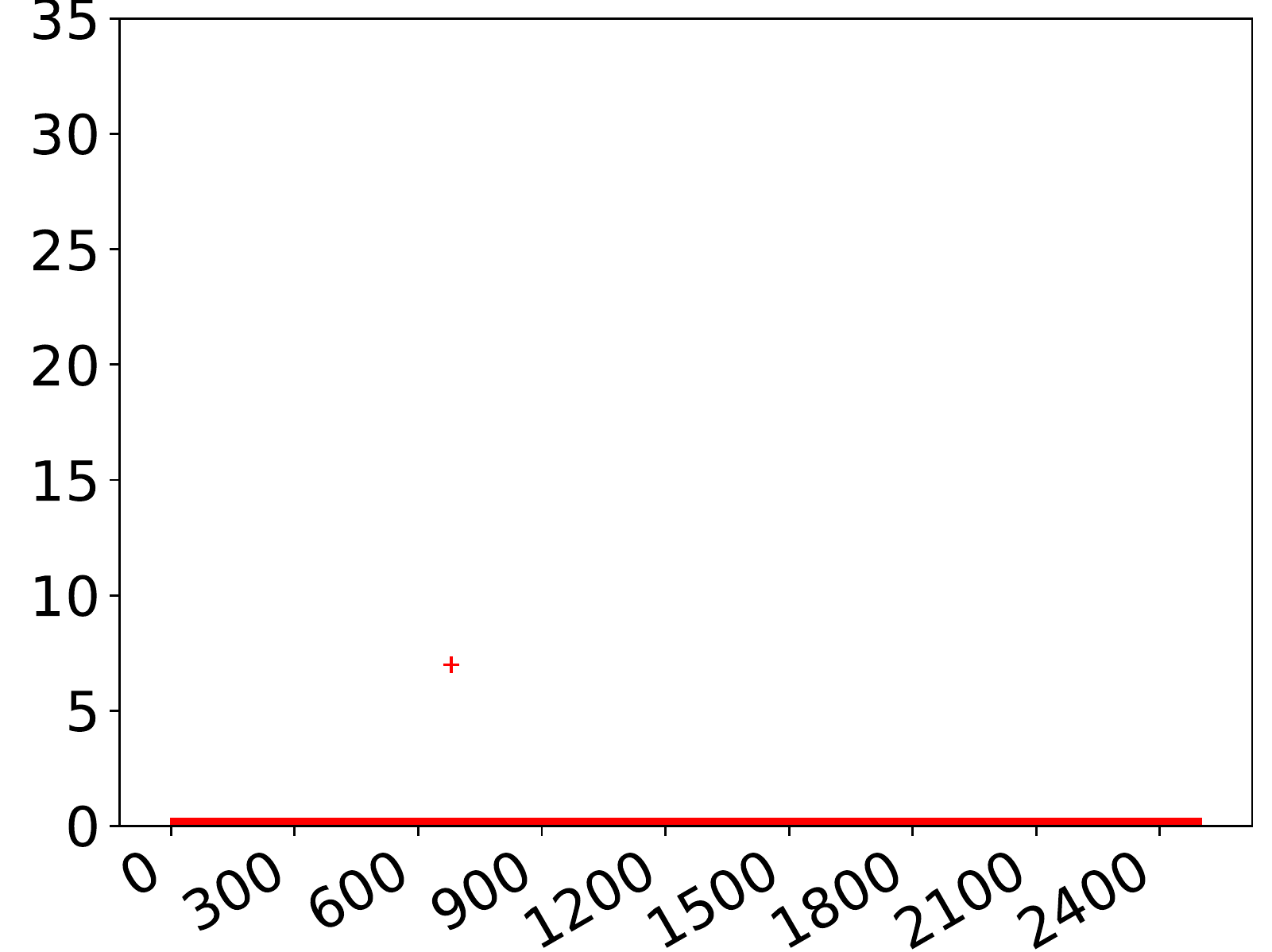}
    \\
    \includegraphics[width=0.22\textwidth]{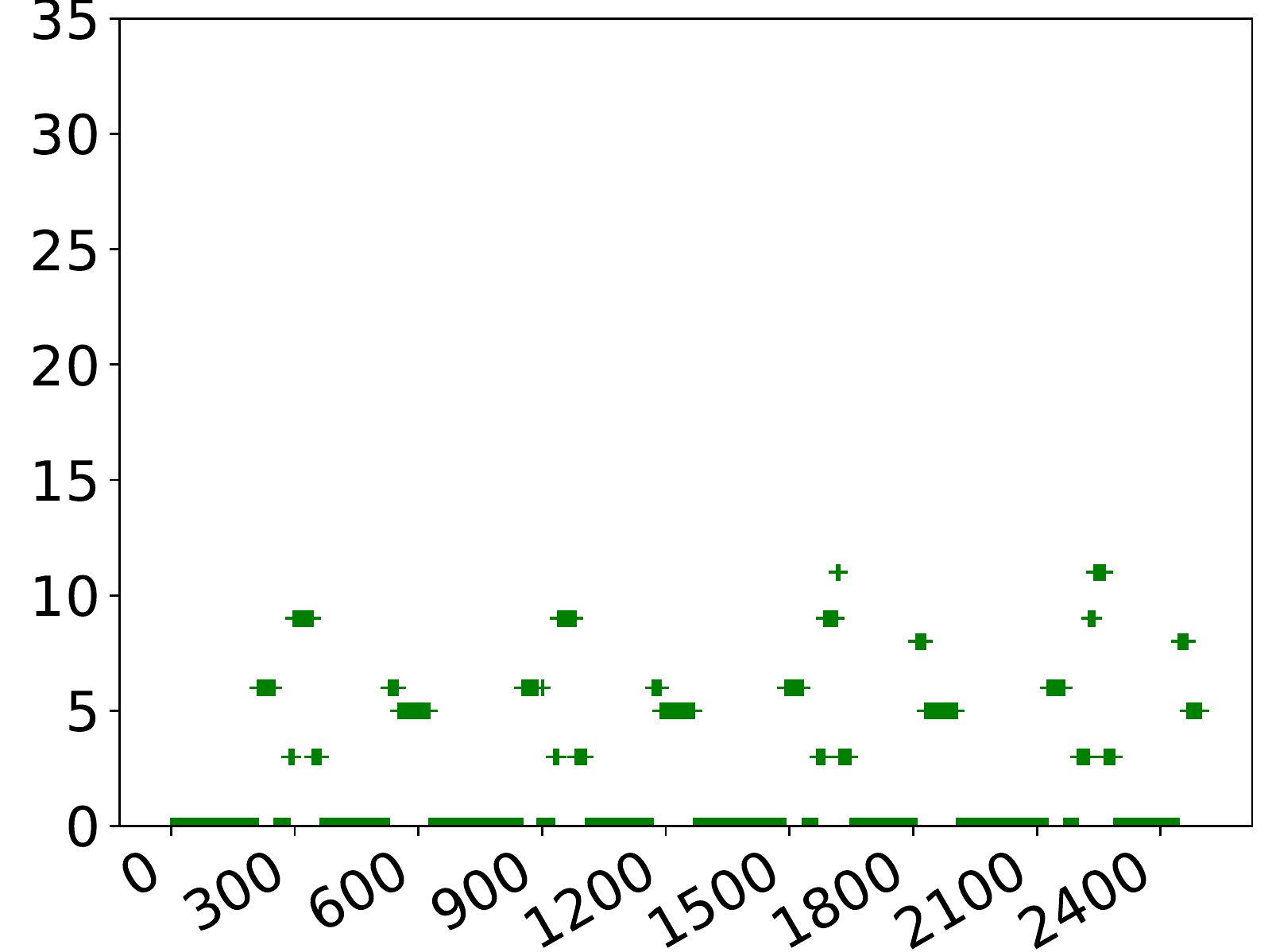}
    ~ 
    \includegraphics[width=0.22\textwidth]{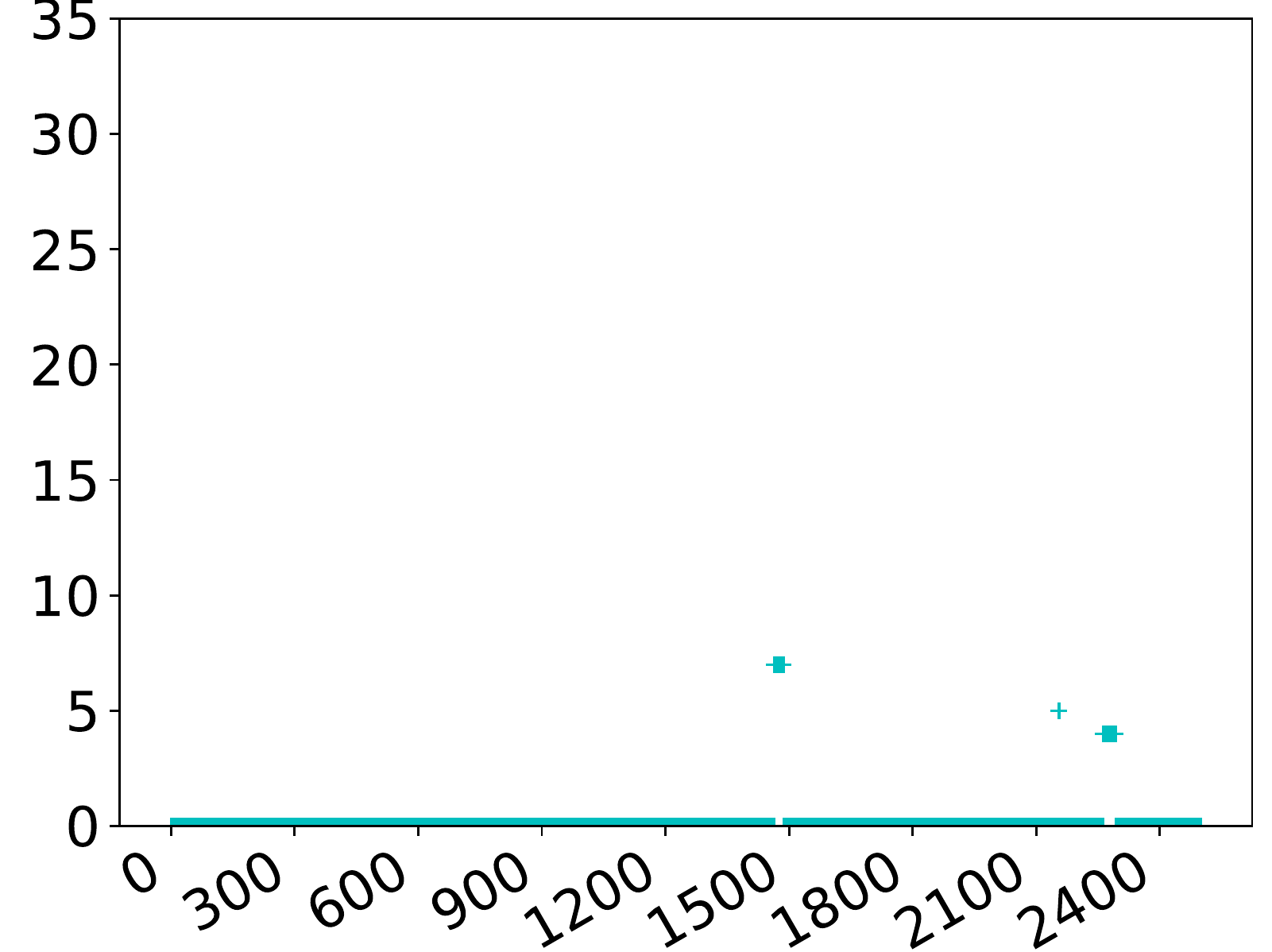}
    \caption{Offset Analysis for the four resistant audio examples by Qin et al.~\cite{qin2019imperceptible}, provided on their website ($y$ axis: edit distance; $x$ axis: added offset in samples).}
    \label{fig:qinRobust}
\end{figure}

\begin{figure}[h]
    \centering
    \includegraphics[width=0.22\textwidth]{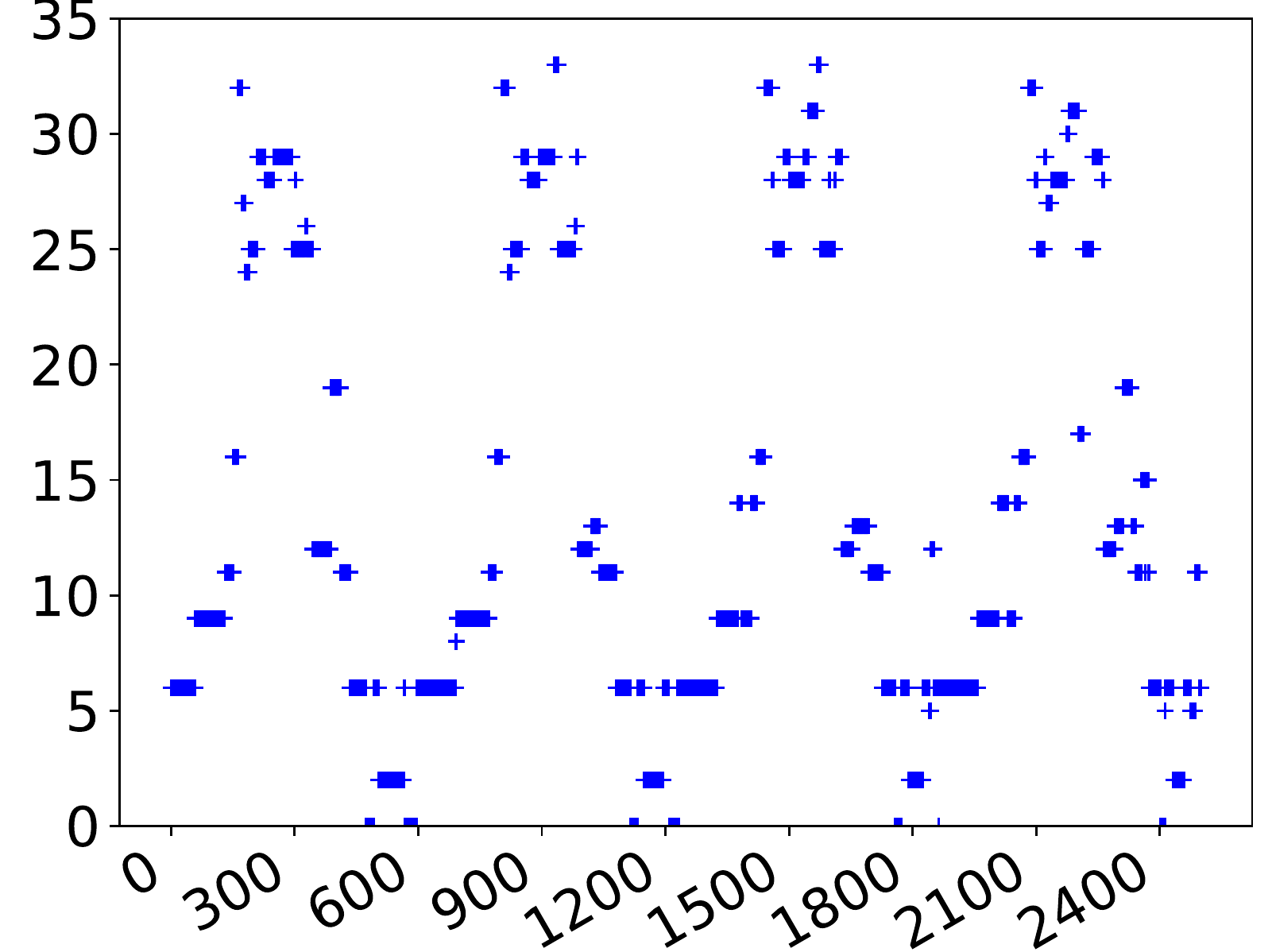}
    ~ 
    \includegraphics[width=0.22\textwidth]{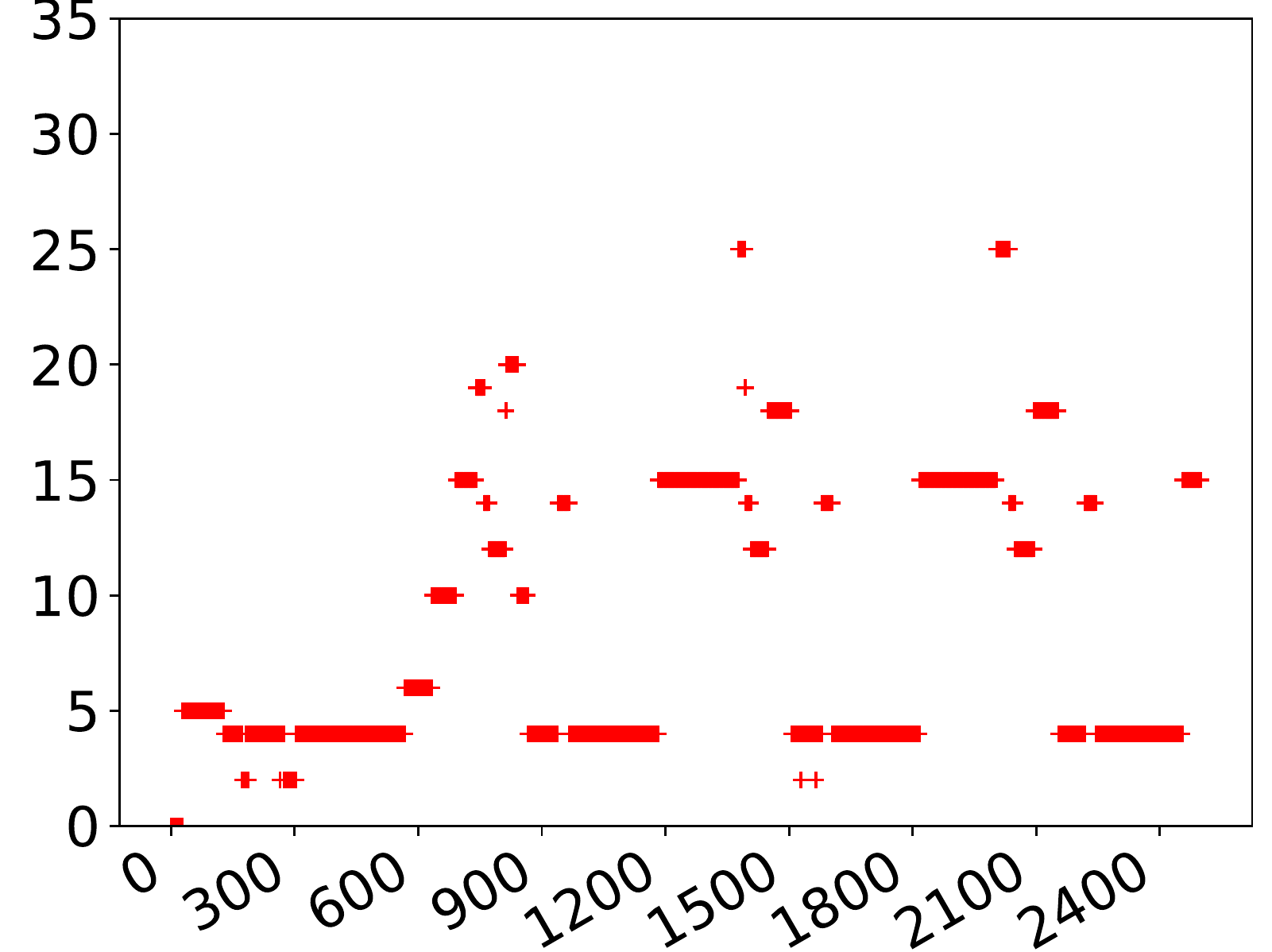}
    \\
    \includegraphics[width=0.22\textwidth]{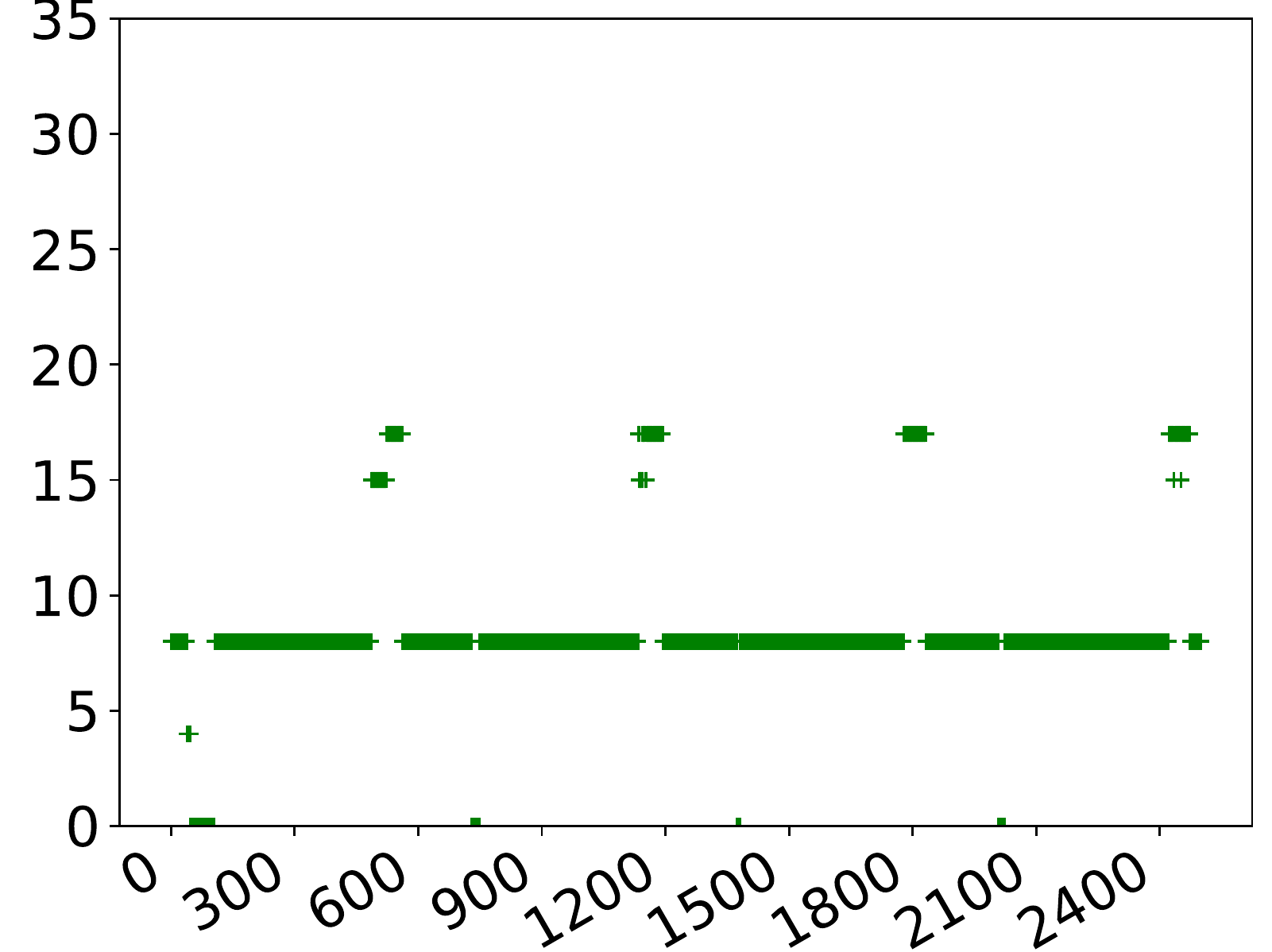}
    ~ 
    \includegraphics[width=0.22\textwidth]{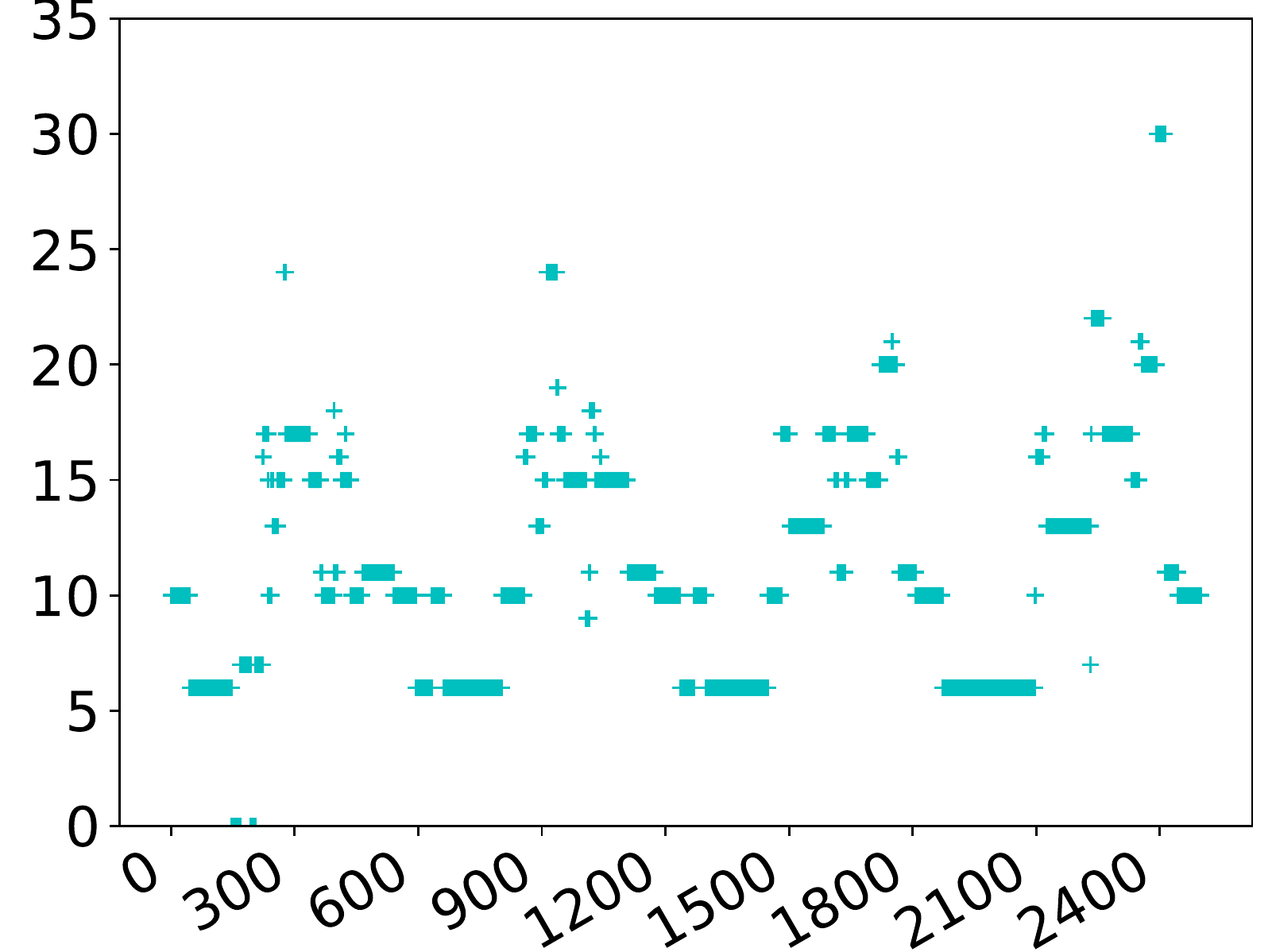}
    \caption{Offset analysis for the four imperceptible and resistant audio examples by Qin et al.~\cite{qin2019imperceptible}, provided on their website ($y$ axis: edit distance; $x$ axis: added offset in samples).}
    \label{fig:qinImperceptible}
\end{figure}

Based on the speech recognition system Lingvo, Qin et al.~\cite{qin2019imperceptible} present a method to generate audio adversarial examples that incorporate a psychoacoustic model and simulated room reverberations.
With those adversarial examples\footnote{All audio files for Qin et al.~are taken from \url{http://cseweb.ucsd.edu/~yaq007/imperceptible-robust-adv.html}, last checked December 17, 2019.}, we perform the same experiment as above and prepend varying lengths of null vector silence to then evaluate how the edit distance changes.
As shown in Figure~\ref{fig:qinRobust} and Figure~\ref{fig:qinImperceptible}, the extent to which the adversarial examples are resistant against changes in offset depends on the generation method. 
For both methods, for the resistant as well as for the resistant and imperceptible adversarial examples, though, we see a certain periodicity in the edit distance.
In contrast to Deepspeech, Lingvo uses a hop size of 512.
Thus, the periodicity in the edit distance is 512 samples, as can be seen from Figure~\ref{fig:qinRobust} and Figure~\ref{fig:qinImperceptible}.

The files used in Figure~\ref{fig:qinRobust} are generated by mainly considering resistance while allowing for the distortion to be perceivable by a human listener.
They incorporate reverberations better, and, hence, implicitly also seem to account for some offset.
This can be deduced from the finding that the edit distances are rather small among all considered offset values.

On the other hand, the files generated with respect to the perceptibility consider room simulations with less importance.
Hence, they are more susceptible to offsets (see Figure~\ref{fig:qinImperceptible}), since the resistance is not the primary objective.
In order to match the input size for Lingvo, \cite{qin2019imperceptible} requires some padding, which is why for most of these examples, there is a non-zero edit distance to the target when no offset is prepended.

\section{Model Definition}\label{sec:problemDefinition}
In the following, we motivate, classify, and explain our model based on \cite{carlini2018audio} according to the taxonomy in \cite{hu2019adversarial}.

\subsection{Motivation for our Model}
As with rotations in the image domain, offsets occur in every real world attack, since no speech recognition system starts listening at the exact moment an adversarial audio signal arrives at the microphone.

While one could conceptualize timing the playback of our adversarial examples such that the speech recognition system receives the adversarial signals at the exact right moment, we think that such attempts are impractical and in vain.
Such attacks would require very precise knowledge about the distance between the loudspeaker used to play the adversarial audio and the microphone used by the speech recognition system.
Deviations from the distance estimate lead to an increased error rate, because the sample rate that is used for recording the audio leads to an offset of multiple samples.
This is due to the slowdown the signals experience when transmitted over the air:
transmitting them over the air happens at a speed of around $340\ms$.
At a sample rate of $16\kHz$ we get an offset of one sample for every $340\ms \cdot (1/16000\sneg) = 0.021\m$ between loudspeaker and microphone, implying that we have an offset of one sample for every $2.1\cm$ of difference in the distance.

If we knew both, the exact moment the speech recognition system starts listening and the distance between the microphone and the loudspeaker, timing the attack would still be challenging since one sample is played every $1/16000\s$, leading to an offset of one sample for every $62\ns$ of delay in playback.

Thus, it is difficult to avoid offsets when attacking speech recognition systems over the air.
Consequently, in order to generate more resistant adversarial examples, we need to take the offset into account.

\subsection{Model Classification}
Our method can be classified as displayed in Table~\ref{tab:taxonomyAccordingToHu} according to the taxonomy in \cite{hu2019adversarial}.
\begin{table}[h]
    \centering
    \caption{Classification of our proposed model according to the taxonomy in \cite{hu2019adversarial}}
    \begin{tabular}{rl}
    \toprule
        \textbf{Adversary Knowledge} & White-Box \\
        \textbf{Adversarial Specificity} & Targeted \\
        \textbf{Target Model} & Deepspeech v0.4.1 \\
        \textbf{Target Object} & Waveform \\
        \textbf{Adversary Method} & Optimization \\
        \textbf{Over-the-Air} & Evaluated on four examples \\
        \textbf{Open Source} & Yes\\
    \bottomrule
    \end{tabular}
    \label{tab:taxonomyAccordingToHu}
\end{table}

\subsection{Algorithm Description}
In this subsection, we describe how we improve the adversarial training process presented in \cite{carlini2018audio} in order to achieve better offset-resistance and, hence, higher resistance against the air gap.
Let $x$ denote the original audio file, and $x_{i-1} = x + \delta_{i-1}$ the adversarial audio file after iteration $i-1$ in the gradient descent calculation from \cite{carlini2018audio}. 
Furthermore, let $R$ denote a random variable distributed uniformly among $\left\{1, \, \ldots, \, 320 \right\}$.
Let $v_1 || v_2$ denote the concatenation of two vectors of audio data $v_1$ and $v_2$.

Our suggested improvement consists of padding the original audio $x$ with silence during the calculation of the adversarial perturbation $\delta$.
More formally, define

\begin{equation*}
    x^0_{i} = \vec{0}^{320-R} || x_{i-1} || \vec{0}^R
\end{equation*}
as the zero padded input for iteration $i$, where we prepend a zero vector of size $320-R$ and append a zero vector of size $R$.
We perform a gradient descent step, and compute the gradient $\nabla_{\delta} L_{\text{CTC}}(x^0_{i}, t, \theta)$ with respect to $\delta$, the adversarial noise.
Here, $t$ denotes the chosen target label.
The weights of the model $\theta$ remain fixed, as is standard when calculating adversarial samples via gradient descent.
We truncate the gradient $\nabla_{\delta} L_{\text{CTC}}(x^0_{i}, t, \theta)$ by removing the first $320-R$ data points and the last $R$ data points.
This removes the gradient update on the zero vector we used for padding.
The resulting vector has the same size as $x$ and $\delta$.
Finally, we apply the gradient to $\delta$ in order to get $\delta_{i}$, calculate $x_{i} = x + \delta_{i}$ and repeat.

\begin{figure}
    \centering
    \includegraphics[width=0.45\textwidth]{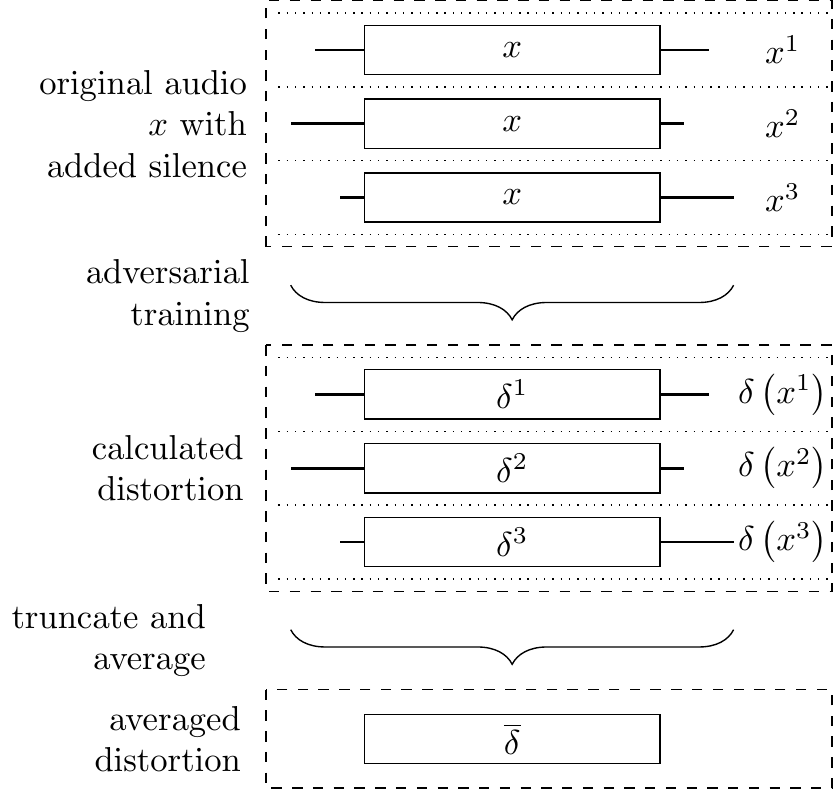}
    \caption{\textit{Batch training.} Parallelizing the generation of adversarial noise using random silence padding:
    first, prepend and append silence to the original audio file $x$ to generate $x^i$, where $i \in \left\{1, \, \ldots, \, B \right\}$ for a batch size of $B$.
    Here, the total silence is fixed to 320, the step size of the MFCC binning operation. 
    Then, for each $x^i$, calculate the gradient with respect to the noise $\delta$, and truncate the result to obtain gradient updates $\delta^i$ of the same size and with correct alignment to $\delta$ and $x$.
    Finally, average all $\delta^i$ to obtain $\overline{\delta}$. 
    According to research in the image domain, averaging perturbations such as padding makes the adversarial example more resistant.
    }
    \label{fig::batchTraining}
\end{figure}

It is possible to parallelize the process described in the section above via batch training.
For a batch size of $B$, we compute $B$ adversarial noise updates as sketched in Figure~\ref{fig::batchTraining}:
the inputs $x^b$ for $b \in \left\{1, \, \ldots, \, B\right\}$ are created by padding $x$ with absolute silence. 
Then, the gradient of $\delta$ is found, truncated to the original size of $\delta$, averaged over all $b$ and added to $x$.

\section{Empirical Evaluation}\label{sec:empiricalEvaluation}
In this section, we evaluate if our proposed method results in the desired resistance properties.
That is to say, we test if for the same level of distortion as in \cite{carlini2018audio}, we get better offset-resistant adversarial examples when fed directly to the speech recognition system. 
Furthermore, we evaluate the edit distances for examples transmitted over-the-air in a realistic room setting and show that our approach indeed improves the adversarial examples' resistance.

\subsection{Number of Iterations}
We first make some theoretical evaluations regarding the training process. 
We find that with relatively few iterations, the adversarial noise is subject to a large variety of different choices for $R$, the length of zero padding.

For $N$ training iterations (with a batch size of $B=1$), the expected number of different padding offsets is
\[
    320 \left ( 1 - \left ( \frac{319}{320} \right ) ^ N \right ).
\]
This can be seen by observing the analog problem of drawing $N$ balls with replacement from a bin of 320 different balls~\cite{stackexchange_balls}.
For $N=1000$ iterations, we can already expect to train on 306 out of 320 possible offsets.

To test if it is possible to generate completely offset-invariant audio with this implementation, we generated adversarial audio based on setting 5 from Table~\ref{tab:settings}: 
a four-second clip of the song ``To The Sky'' by Owl City used by Yuan et al.\footnote{See \url{https://sites.google.com/view/commandersong/}, last checked December 12, 2019.}~\cite{yuan2018commandersong} with the target 'open the door' and a maximum allowed distortion of $70.0\dB$.

In this setup, even for 1000 training iterations, the edit distance is exactly zero for every offset. 
This implies that the audio stays invariant to offsets it has not been trained on in the last iterations, as there is at least one offset distance $R$ that the audio has not been trained on for at least $320-1$ iterations.
The offset-invariant adversarial audio example as well as an adversarial audio example that was not optimized for offset invariance are provided online. 

\begin{figure}[h]
    \centering
    \includegraphics[width=0.5\textwidth]{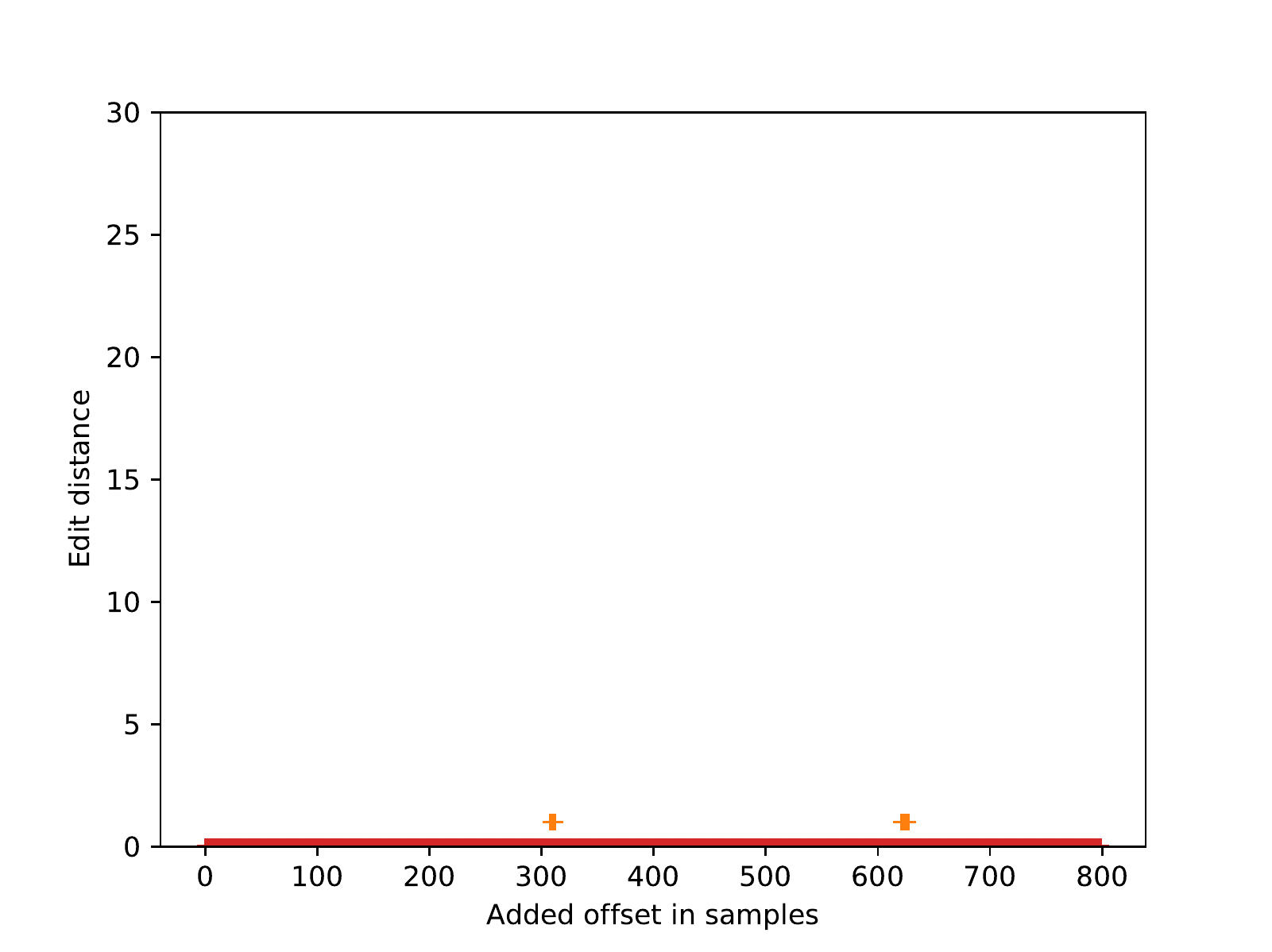}
    \caption{
        Offset analysis for four adversarial audio files (same original - target combination as in Figure~\ref{fig:targetedCarlifiniOffsetVariation10Samples}) that were generated with the offset training ($y$ axis: edit distance; $x$ axis: added offset in samples). Axis limits chosen as in Figure~\ref{fig:targetedCarlifiniOffsetVariation10Samples} to improve comparability.
    }
    \label{fig:offset_robust_4}
\end{figure}

\subsection{Theoretical Offset-Resistance}
Changing the offset in every iteration results in almost complete invariance against offsets for the combination of audio and adversarial label used.
This can be seen from Figure~\ref{fig:offset_robust_4}.
Here, the same examples as in Figure~\ref{fig:targetedCarlifiniOffsetVariation10Samples} are calculated with our adapted algorithm (one of the examples is displayed in Figure~\ref{fig:teaser}).
There is only one example, plotted in orange, that is not resistant against every offset value as it differs in an edit distance of one.

In comparison to the approach in \cite{carlini2018audio}, we could achieve a nearly perfect offset-resistance for all considered examples.

\begin{table}[h]
    \centering
    \caption{Comparison of the edit distances for over-the-air recorded original examples, averaged over twenty recordings per file (characters for original label)}
    \begin{tabular}{ccc}
    \toprule
        Setting & Edit Distance & Characters \\
    \midrule
        1 & 15.7 & 28 \\
        2 & 15.2 & 60 \\
        3 & 15.4 & 29 \\
        4 & 24.0 & 51 \\
    \bottomrule
    \end{tabular}
    \label{tab:comparisonOverTheAirOriginal}
\end{table}

\subsection{Evaluation Over-the-Air}
An evaluation (see Table~\ref{tab:comparisonOverTheAir}) in an over-the-air setting suggests that including offset in the training improves the resistance in general.

Again, we have analysed the examples from settings 1-4 (see Table~\ref{tab:settings}).
In order to see how well Deepspeech v0.4.1 can cope with realistic recordings of audio files, we have played and recorded the original files with the following setup.
Each file was played with a Logitech X530 speaker and recorded by a Rode NT-USB microphone. 
The speakers and the microphone were $10\cm$ apart in a regular home setting (fully furnished, no certain noise dampening, usual reverberations).
When analysed with Deepspeech, we get the edit distances displayed in Table~\ref{tab:comparisonOverTheAirOriginal}, which are averaged over twenty recordings per file.
For comparison, we have also included the length of the phrases measured in characters (including spaces and apostrophes).
As one can see, in general, Deepspeech fails to transcribe the noisy audio files correctly\footnote{This seems to be a known issue, see \url{https://github.com/mozilla/DeepSpeech/issues/2077}.}.

In a second step, we compare the over-the-air resistance of adversarial examples constructed based on \cite{carlini2017adversarial} and our approach, see Table~\ref{tab:comparisonOverTheAir}.
As a measure for success, we have chosen the edit distance.
Again, we have averaged the edit distance between the target and the transcribed label over twenty recordings per audio file. 
Note that the distances have not been normalized to the length of the target label or original label.

In order to improve comparability, the adversarial noise in the non-offset training, as introduced in \cite{carlini2017adversarial}, and offset training, as presented in this work, has been generated with respect to the same upper limit of $66.02\dB$ (for a raw signal range of $\pm 1000$ in pulse-code modulation encoding).
In all four settings, the offset-training outperforms the non-offset-training significantly.
For most of the settings, our results seem to be comparable to the edit distances for the original files when recorded with the same setup and evaluated with Deepspeech (see Table~\ref{tab:comparisonOverTheAirOriginal}).
We speculate that including this kind of training in more recent attacks and more noise robust models, e.g.~\cite{qin2019imperceptible} (see Section~\ref{ssec:qin}), will lead to even better edit distances.
\begin{table}[h]
    \centering
    \caption{Comparison of the edit distances for over-the-air recorded adversarial examples, averaged over twenty recordings per file (OT: offset-training; characters for target label)}
    \begin{tabular}{cccc}
    \toprule
        Setting & Without OT & With OT & Characters \\
    \midrule
        1 & 28.0 & 17.6 & 29 \\
        2 & 35.3 & 29.6 & 51 \\
        3 & 19.2 & 18.9 & 22 \\
        4 & 30.3 & 24.7 & 50 \\
    \bottomrule
    \end{tabular}
    \label{tab:comparisonOverTheAir}
\end{table}

We thus conclude that including knowledge on the preprocessing enhances the resistance of adversarial examples when transmitted over-the-air.

\section{Conclusion}\label{sec:conclusion}
Current methods for generating adversarial examples are not yet suited to reliably bridge the air gap.
We identify one necessary condition for air gap resistance: invariance to the offset.
This is necessary since the effectiveness of adversarial examples must not depend on the exact alignment of the MFCC-preprocessing binning during training and testing. 
This is unattainable in real-world scenarios, where there will always be random padding due to environmental factors.
Thus, adversarial training needs to be enhanced by random padding of the input and by finding adversarial noise which is resistant to such random padding.
We design such a training loop and show that it results in offset-invariant adversarial examples and also improves resistance in the over-the-air setting.
We suggest to incorporate this training method into future approaches for generating adversarial examples targeting to bridge the air gap.

\bibliographystyle{ACM-Reference-Format}
\bibliography{audioBib}



\end{document}